\documentclass[twocolumn, 10pt]{article}
\usepackage{amsmath,amssymb}
\usepackage{graphicx}
\usepackage{float}
\usepackage{diagbox}
\usepackage[justification=centering]{caption}
\usepackage{titlesec}
\usepackage{textcomp}
\usepackage{cite}
\usepackage{float}
\usepackage{lineno}
\usepackage{hyperref}
\usepackage{subcaption} 
\usepackage[export]{adjustbox}
\usepackage{etoolbox}
\usepackage{centernot}

\makeatletter
\def\affiliation#1{\gdef\@affiliation{#1}}
\def\abstract#1{\gdef\@abstract{#1}}
\def\graphabst#1{\gdef\@graphabst{#1}}
\def\keywords#1{\gdef\@keywords{#1}}
\def\corresp#1{\gdef\@corresp{#1}}

\newcommand{\MakeTitle}{
  \newpage
  \null
  \vskip 2em%
  \begin{center}%
  \Large \@title\par
  \vskip 1em%
  \large \@author
  \end{center}
  \noindent\@affiliation\par
  \vskip 1em%
  \noindent\@corresp\par
  \vskip 1em%
  \noindent\@abstract\par
  \vskip 1em%
  \noindent\@keywords\par
}
\makeatother

\makeatletter
\patchcmd{\@maketitle}{\raggedright}{\centering}{}{}
\patchcmd{\@maketitle}{\raggedright}{\centering}{}{}
\makeatother

\newcommand*{\TitleFont}{%
      \usefont{\encodingdefault}{\rmdefault}{}{n}%
      \fontsize{18}{12}%
      \selectfont}

\titleformat{\section}
  {\normalfont\fontsize{10}{11}\bfseries}{\thesection.}{2pt}{}
  \titlespacing*{\section}{0pt}{12pt}{6pt}

\titleformat{\subsection}
  {\normalfont\fontsize{10}{10}\bfseries}{\thesubsection.}{2pt}{}
  \titlespacing*{\subsection}{0pt}{6pt}{0pt}
  
\titleformat{\subsubsection}
  {\normalfont\fontsize{10}{10}\bfseries}{\thesubsubsection.}{2pt}{}
  \titlespacing*{\subsubsection}{0pt}{6pt}{0pt}
  
\normalsize
\title{\TitleFont Demystifying the nonlocality problem in Aharonov-Bohm effect}

\author{Kolahal Bhattacharya}

\affiliation{
\begin{center}
Manipal Academy of Higher Education, Manipal, KA-576104, India\\
Homi Bhabha Centre for Science Education, TIFR, Mumbai - 400088, India
\end{center}
}

\corresp{$^{\ast }$kolahal@hbcse.tifr.res.in
}
\abstract{\textbf{Abstract}:
In this paper, we present a novel semi-classical theory of the electrostatic and magnetostatic fields and explain the nonlocality problem in the context of the Aharonov-Bohm effect~\cite{aharonov1959significance}. Specifically, we show that the electrostatic and the magnetostatic fields possess a quantum nature that manifests if certain conditions are met. 
In particular, the wave amplitudes of the fields are seen to exist even in the regions where the classical fields vanish and they operate on the electron wave functions locally as unitary phases. This formulation also sheds light on the quantisation of electric charges and magnetic flux.}


\begin{document}
\onecolumn
\MakeTitle
\section{Introduction}
The Aharonov-Bohm effect~\cite{aharonov1959significance} asserts that the charged particles, e.g. electrons, can be affected by the classical potential in a region where the classical field is zero. When it was first conceptualised, it implied that the potentials were more fundamental quantities than the fields in the quantum domain, in spite of the fact that the former entities are not unique. This raised considerable debates about the type of interaction (local or non-local?) of the electrons with the classical fields which vanish in the regions accessible to the electrons. These discussions are alive even in recent times~\cite{vaidman2012role,aharonov2016nonlocality,kang2017proposal}. There is another issue. Though the magnetic Aharonov-Bohm effect has been experimentally confirmed~\cite{tonomura1986evidence}, the electrostatic Aharonov-Bohm effect has not been experimentally observed yet~\cite{matteucci1985new, washburn1987normal, van1998magneto}, since the effect of the classical force could not be completely removed. This led to suspicion from the scientific community~\cite{walstad2010critical} that this effect may not exist. It remains a very controversial issue even today. Some recent works suggest new experimental techniques to test the electrostatic Aharonov-Bohm effect~\cite{kim2018electric, bachlechner2020proposal}. Therefore, we can safely say that the complete understanding of Aharonov-Bohm effect is yet to be achieved.

In this work, we take a fresh perspective to the problem. In classical physics, an electron is characterised by a charge $-e$ and it can feel a force $|\mathcal{F}|=e\mathcal{E}$. But in quantum mechanics, the same electron is characterised by a wave function $\psi_e$. So, perhaps it is unwise to expect that the classical prescription about how an electron feels a field will hold true. We must use some equivalent to the wave functions (or operators) of the fields to understand the transformation of the electron states when it is subjected to a static field.

Such a quantum theoretical description is provided by the quantum electrodynamics which is an elegant relativistic quantum field theory of the electromagnetic fields. In this formalism, one can 
recover the Maxwell's equations of electrodynamics by performing Euler-Lagrange variation of the electromagnetic Lagrangian, with respect to the four potential $A^\mu=(\phi,{\bf A})$. In Coulomb gauge, the vector potential ${\bf A}$ satisfies the wave equation and has plane wave solutions of the form $Exp[i({\bf k}\cdot{\bf r}-\omega_kt)]$ for the angular frequency $\omega_k$ and the wave vector $\bf{k}$. The field ${\bf A}$ can be quantised by imposing periodic boundary conditions over a volume $V$. This enables one to express ${\bf A}$ (and hence, physical electromagnetic fields) in this volume as a Fourier sum over all the discrete independent plane wave modes with the corresponding polarisation states. The Hamiltonian of this system is the sum of Hamiltonian of all such modes, which individually can be expressed as sum of the squares of the coordinates and the squares of conjugate momenta. One can, therefore, treat electromagnetic field as an ensemble of quantum harmonic oscillators. Thus, quantisation of the electromagnetic fields can be achieved by canonical quantisation of these modes. The resulting electric and magnetic field operators are expressed as: 

\begin{align}\label{e1}
{\hat{\bf E}}&=\sum_{\alpha=1,2}\sum_{k}\sqrt{\frac{2\pi\hbar\omega_k}{V}}\left[a_{k\alpha}{\bf e}^{i({\bf k}\cdot{\bf r}-\omega_kt)} - a_{k\alpha}^\dagger {\bf e}^{-i({\bf k}\cdot{\bf r}-\omega_kt)} \right]\hat{e}_\alpha({k})
    \nonumber \\
{\hat{\bf B}}&=\sum_{\alpha=1,2}\sum_{k}\sqrt{\frac{2\pi\hbar\omega_k}{V}}\left[a_{k\alpha}{\bf e}^{i({\bf k}\cdot{\bf r}-\omega_kt)} - a_{k\alpha}^\dagger {\bf e}^{-i({\bf k}\cdot{\bf r}-\omega_kt)} \right](\hat{k}\times\hat{e}_\alpha({k}))
\end{align}
Does this theory apply to the situation, where only the electric field or only the magnetic field is present, as happens in the case of the Aharonov-Bohm effect? We notice that the interpretation of the electromagnetic field as an ensemble of the harmonic oscillators is possible, only if we add the contributions from both electric and magnetic fields. In the absence of any one of them, we cannot express the Hamiltonian as the Hamiltonian of the simple harmonic oscillator. 

In general, if we have a time-varying electric field, that automatically implies the existence of a rotational magnetic field and vice versa. Therefore, Eq.\eqref{e1} is directly applicable in this case. For static fields, it is not that apparent. In fact, it may appear a little counter-intuitive that even static fields can also be represented as a sum over the travelling wave states, as in Eq.\eqref{e1}. After all, there is no question of frequency in electrostatic or magnetostatic fields. However, there is no contradiction because the coefficients of the plane waves in the sum can be selected in such a way that one can effectively construct a static field. 

It will probably be, however, physically more intuitive to have a theory of the electrostatic or magnetostatic fields that does not require any component travelling waves. Photon description of the fields in quantum electrodynamics (as in Eq.\eqref{e1}) is Lorentz-covariant and has the information about polarisation. However, such considerations do not come into the discussion when talking about classical static fields. Hence, it will be no surprise if the an alternative quantum description of the fields do not have these features.

We approach the problem using a novel variational principle in electrostatics $\delta\int\mathcal{E}\ ds=0$, in the context of a pedagogical problem (method of images)~\cite{bhattacharya2013novel}. In this principle, the line integral must be performed along a curve $C$ which is always superimposed with the local direction of the electrostatic field. Of course, the value of the integral $\int_{A}^{B}\vec{\mathcal E}\cdot\ d{\vec{s}}$ is stationary along any curve connecting the two fixed points $A$ and $B$. However, if we integrate along the curve $C$, then one can verify that the Euler-Lagrange equation for this principle is consistent with the paths for which we have $\nabla\times \mathcal{\vec{E}}={\bf 0}$. This is the significance of explicitly stating the principle $\delta\int\mathcal{E}\ ds=0$.

There is no way one can miss the strong analogy of this principle to the Fermat's principle in optics, $\delta\int{n\ ds}=0$, where $n$ denotes the refractive index of the medium. It is possible to develop a Hamiltonian formulation of geometrical optics, based on the Fermat's principle~\cite{buchdahl1993introduction}. Light ray follows the paths obtained by the Euler-Lagrange variation of the Fermat's principle. A quantum theory of light rays was constructed by Gloge et al.~\cite{gloge1969formal}, based on this Hamiltonian formulation of optics. In this work, the probability wave amplitudes of the time-independent reduced wave equation were called the `wave functions' of the light rays. It was also argued that geometrical optics emerges in the limit $\lambda\rightarrow0$ where $\lambda$ denotes the wavelength of the light. This suggests that a similar analytical treatment on $\delta\int\mathcal{E}\ ds=0$ should be possible and this should lead to the corresponding time-independent wave equation of the electrostatic field (similar to the reduced wave equation (Eq.(7)) in~\cite{gloge1969formal}), whose solutions may describe the wave amplitudes of the electrostatic fields.

Now, there is nothing special about the electrostatic fields. One can easily check that a similar variational principle $\delta\int \mathcal{F}\ ds=0$ can be conceived for any curl-free field $\vec {\mathcal{F}}$, when the integral is performed along a curve that is superimposed with the local direction of $\vec{\mathcal{F}}$. 
Thus, this will also hold true in somewhat special situations in magnetostatics, e.g. (a) $\delta\int\mathcal{H}\ ds=0$ when free current density $\vec {J}_{free}={\bf 0}$ (from Ampere's law $\nabla\times\vec {\mathcal{H}}\equiv\vec {J}_{free}={\bf 0}$), and (b) $\delta\int \mathcal{A}\ ds=0$ for the electromagnetic vector potential $\vec{\mathcal A}$, if the magnetic field $\vec{\mathcal B}$ itself is zero in a region, i.e. $\nabla\times\vec{\mathcal{A}}\equiv\vec {\mathcal{B}}={\bf 0}$. In all these cases, it is understood that the integral is evaluated along a curve always superimposed with the local direction of the corresponding fields. Therefore, it is apparent that in all these cases, one can develop the Lagrangian or Hamiltonian formulations, in the same way Hamiltonian formulation of optics is developed. 

This paper is organised in the following manner. In the next section~\ref{Sec2}, we develop the semi-classical theory of  the electrostatic and magnetostatic fields. In this section, we also introduce concepts corresponding to Bohr-Sommerfeld quantisation rule. In the following section~\ref{Sec3:AB}, we apply the concepts to resolve the nonlocality problem in the Aharonov-Bohm effect. In the final section~\ref{Sec4:RWE}, we apply the semi-classical theory to several other questions, including the quantisation of electric charges. It is shown that in general, electric charges must be a rational multiple of the elementary electric charge and its conception does not require the existence of magnetic monopoles.

\section{Hamiltonian formulation and quantization procedure}\label{Sec2}
In this section, we shall derive the scalar wave equation of the electrostatic fields [Eq.\eqref{eq7} in the following]. This can also be derived using the method adopted in~\cite{gloge1969formal}). 
\subsection{Electrostatic case}
\subsubsection{Hamiltonian formulation}
In terms of a dimensionless stepping parameter $a$ the electrostatic variational principle can be expressed as
\begin{equation}\label{eqELag}
    \delta\int\mathcal{E}\left(\frac{ds}{da}\right)\ da=0
\end{equation}
-where Eq.\eqref{eqELag} corresponds to the Lagrangian $\mathcal{E}\left(\frac{ds}{da}\right)$ which has the dimension of electrostatic potential. It can be written as:
\begin{equation*}
    L(x,y,z,x',y',z',a)=\mathcal{E}(x,y,z)\sqrt{(x'^2+y'^2+z'^2)}
\end{equation*}
-where $x'=\frac{dx}{da}$ etc. Hence, the conjugate momenta can be calculated as:
\begin{align}\label{eqE-conjmom}
    p_{x}&=\frac{\partial L}{\partial x'}=\mathcal{E}\frac{x'}{\sqrt{x'^2+y'^2+z'^2}}=\mathcal{E}\frac{dx}{ds}=\mathcal{E}_x\nonumber\\
    p_{y}&=\frac{\partial L}{\partial y'}=\mathcal{E}\frac{y'}{\sqrt{x'^2+y'^2+z'^2}}=\mathcal{E}\frac{dy}{ds}=\mathcal{E}_y\nonumber\\
    p_{z}&=\frac{\partial L}{\partial z'}=\mathcal{E}\frac{z'}{\sqrt{x'^2+y'^2+z'^2}}=\mathcal{E}\frac{dz}{ds}=\mathcal{E}_z
\end{align}
The Hamiltonian can be evaluated as follows:
\begin{align*}
    H(x,y,z,p_x,p_y,p_z,a)&=x'p_{x'}+ y'p_{y'}+z'p_{z'}-L\nonumber\\
    &=\mathcal{E}\frac{x'^2}{\sqrt{x'^2+y'^2+z'^2}}+\mathcal{E}\frac{y'^2}{\sqrt{x'^2+y'^2+z'^2}}+\mathcal{E}\frac{z'^2}{\sqrt{x'^2+y'^2+z'^2}}-L\nonumber\\
    &=\mathcal{E}\sqrt{x'^2+y'^2+z'^2}-\mathcal{E}\sqrt{x'^2+y'^2+z'^2}\nonumber\\
    &=0
\end{align*}
In the Hamiltonian formulation, the electrostatic configuration of a system can be represented by a phase space $(x,y,z,\mathcal{E}_x,\mathcal{E}_y,\mathcal{E}_z)$. A field point in the real field is represented by a unique point in this phase space. Classically, we see a continuum of points each one of which has $H=0$.

\subsubsection{Transition to quantum domain}
The transition to classical to quantum theory requires that the momenta $p_{x}, p_{y}, p_{z}$ are treated as operators:
\begin{align}\label{eqE-mom}
    p_{x'}\rightarrow\hat{p}_x=-i\frac{\gamma}{2\pi}\frac{\partial}{\partial x}\nonumber\\
    p_{y'}\rightarrow\hat{p}_y=-i\frac{\gamma}{2\pi}\frac{\partial}{\partial y}\nonumber\\
    p_{z'}\rightarrow\hat{p}_z=-i\frac{\gamma}{2\pi}\frac{\partial}{\partial z}
\end{align}
operating on a wave amplitude $\psi_E$ of the electrostatic field. In Eq.\eqref{eqE-mom}, $\gamma$ is independent of the spatial coordinates $(x,y,z)$ and has the dimension of the electrostatic potential. It plays the role of Planck's constant in this equation. That is, the quantum aspect of the electrostatic field is only visible in the limit when $\gamma$ cannot be negligible with respect to the potentials in the problem ($\gamma\centernot\rightarrow{0}$).
On the other hand, in the limit $\gamma\rightarrow0$, we have classical electrostatics, in the same way the limit $\hbar \rightarrow0$ denotes the transition from quantum mechanics to classical mechanics and the limit $\lambda\rightarrow0$ represents the transition from wave optics to geometrical optics. So, the theory we could have for electrostatics can probably be called wave electrostatics.

We define $\psi_E$ through the eigenvalue equation $\hat{p}_x\psi_E= \mathcal{E}_x\psi_E$ etc.\footnotemark[1] Therefore, $\hat{p}_x^2\psi_E=\mathcal{E}_x\hat{p}_x\psi_E=\mathcal{E}_x^2\psi_E$ (here $\hat{p}_x$ does not operate on $\mathcal{E}_x$, which is an eigenvalue and not a state). This should be equal to:
\footnotetext[1]{Then, $\hat{\vec{p}}\psi_E=-i\frac{\gamma}{2\pi}\nabla\psi_E=(\hat{x}\hat{p}_x+ \hat{y}\hat{p}_y+\hat{z}\hat{p}_z)\psi_E=(\hat{x}\mathcal{E}_x+\hat{y}\mathcal{E}_y+\hat{z}\mathcal{E}_z)\psi_E=\vec{\mathcal{E}}\psi_E$.}
\begin{align*}
    \hat{p}_x^2\psi_E &=-\bar\gamma^2\frac{\partial^2}{\partial x^2}\psi_E
\end{align*}
where $\bar\gamma=\frac{\gamma}{2\pi}$. Similar equations hold true for the other coordinates. Thus, summing over all the coordinates, we find:
\begin{align}\label{eq7}
\mathcal{E}^2\psi_E=-\bar\gamma^2\nabla^2&\psi_E\nonumber\\
    \implies \bar\gamma^2\nabla^2\psi_E+\mathcal{E}^2\psi_E&=0
\end{align}
This equation is the scalar wave equation of the wave amplitude of the electrostatic field. The assertion that $\gamma\centernot\rightarrow0$ in a given problem signifies that the continuum structure of points in the phase space $(x, \hat{p}_x=\mathcal{E }_x,y,\hat{p}_y=\mathcal{E}_y,z,\hat{p}_z=\mathcal{E}_z )$ does not hold any longer. Hence, the electrostatic configuration of a system can be best determined up to a minimum value $\gamma$ that represents the area of an elementary cell in the phase space. We can easily check that $\hat{x}$ and its conjugate momentum $\hat{p }_x$ do not commute.

In this semi-classical model of the electrostatic field, one can find the allowed quantised states $\psi_E$ of a system by requiring that these are the states which follow the Bohr-Sommerfeld quantisation condition in this context:
\begin{equation}\label{EqBohrSommerfeld}
    \oint\mathcal{E}_x\ dx=N_x\gamma,\hspace{0.5 cm}
    \oint\mathcal{E}_y\ dy=N_y\gamma,\hspace{0.5 cm}
    \oint\mathcal{E}_z\ dz=N_z\gamma\hspace{0.5 cm}
\implies\oint\vec{\mathcal{E}}\cdot{d{\vec{ s}}}= (N_x+N_y+N_z)\gamma
\end{equation}
-where $N_{x,y,z}\in[1,2...]$. The last integral shown in Eq.\eqref{EqBohrSommerfeld}, $\oint\vec{\mathcal{E}} \cdot{d{\vec{s}}}$, classically evaluates to zero since $\gamma\rightarrow0$. 

Eq.\eqref{eq7} shows that the wave amplitude $\psi_E$ of the electrostatic field can be thought of as the scalar wave or disturbance that represents the electrostatic field somewhat similar to the Huygen's wavelets used in wave optics. Let us take an example: a point charge is placed at the origin of a frame. The charge is at rest and the electrostatic field drops according to inverse square law in this frame. Now, $ \mathcal{E}\rightarrow\infty$ as $r\rightarrow0$. So, we see that Eq.\eqref{eq7} dictates that $\psi_E\equiv0$ at the location of the source charge. So, the waves representing electrostatic field are distributed radially in all directions with a common concurrent node at the origin. Now, let us take the example of two static point charges. In this case, we can visualise $\psi_E$ as the waves along the electrostatic field lines between the two charges, in such a way that nodes always occur at the locations of both the charges (this is somewhat like the vibrations in a non-stretchable string fixed at both the ends). 


In a region where the source charge density $\rho=0$, a simple solution to Eq.\eqref{eq7} is given by $\psi_E({ \bf r})=\frac{1}{\sqrt{2\pi}}{\bf e}^{\frac{i}{\bar \gamma}\int\mathcal{E}\ ds}$. Absorbing the constant coefficient, $\psi_E$ can be expressed as ${\bf e}^{\frac {i}{\bar\gamma}\Phi({\bf r})}$, where $\Phi$ denotes the electrostatic potential at ${\bf r}$. This is the form of the Huygen's planar wavelet for the electrostatic field under semi-classical conditions. It can mediate the local action of the electrostatic field. Since it has the form of a phase or plane wave, it is not normalisable. However,
we can find a general solution $\psi_E'=\int a(\mathcal{E} ){\bf e}^{\frac{i}{\bar\gamma}\Phi({\bf r})}d\mathcal{E}$ to Eq.\eqref{eq7} with non-zero $\mathcal{E}$ by superposing several such plane waves; this resulting wave packet $\psi_E'$ can be normalised. It could denote the virtual photons of the electrostatic field which have never been directly observed\footnotemark[5].
\footnotetext[5]{Aharonov and Bohm, in their original paper~\cite{aharonov1959significance}, used the ``gauge transformed'' electron wave function: $\psi_e \rightarrow\psi_e'={\bf e}^{\frac{i}{\hbar}e\Phi t}\psi_e $. However, with normalised wave packets of electrostatic field, the electron wave function would transform as: $ \psi_e\rightarrow\psi_e''=\left(\int a(\mathcal E){\bf e}^{\frac {i}{\hbar}e\Phi t}d\mathcal{E}\right)\psi_e$ (with the knowledge of $\bar\gamma$). Now, in case of the electrostatic Aharonov-Bohm effect, $\mathcal{E}=0$ and thus, $a(\mathcal E)$ is constant. So, it makes more sense to use $\psi_e'$ instead of $\psi_e''$. 

In section 2 of the paper~\cite{aharonov1959significance}, Aharonov and Bohm showed that $\psi_e'$ is the solution of the Schr$\rm\ddot{o}$dinger's equation with a Hamiltonian $H_0+V(t)$, where $\psi_e$ is the solution of the Schr$\rm \ddot{o}$dinger's equation with Hamiltonian $H_{0}$. Next, they superposed $\psi_e^{1'}$ and $\psi_e^{2'}$ from the two separate streams of electrons.

We can check that $\psi_e''$ is a solution of the Schr$\rm \ddot{o}$dinger's equation with a Hamiltonian $H_0+V(t)$:
\begin{equation*}
i\hbar\frac{\partial\psi_e''}{\partial t}=i\hbar\left(\int a(\mathcal{E}){\bf e}^{\frac{i}{\hbar}e\Phi t}d\mathcal{E} \right)\frac{\partial\psi_e}{\partial t}+i\hbar\left(\int a(\mathcal{E}){\bf e}^{\frac{i}{\hbar}e\Phi t}d\mathcal{E} \right)\left(\frac{i}{\hbar}e\Phi\right)\psi_e=\left(i\hbar\frac{\partial}{\partial t}-e\Phi\right)\psi_e''=\left (H_0+V(t)\right)\psi_e''
\end{equation*}
But this is relevant only if $\mathcal E\neq0$. Under such circumstances, the electrons from the two separate streams must be superposed as:
\begin{equation*}
    \Psi_e=\left(\int a_1(\mathcal E){\bf e}^{\frac {i}{\hbar}e\Phi_1 t}d\mathcal{E}\right)\psi_e^{1}+\left(\int a_2(\mathcal E){\bf e}^{\frac {i}{\hbar}e\Phi_2 t}d\mathcal{E}\right)\psi_e^{2}
\end{equation*}
This is correct when the electric field cannot be reduced exactly to zero. But if the field $\mathcal{E}=0$, then we must use the non-normalisable wave amplitude $\psi_E=e^{\frac{i}{\hbar}e\Phi t}$.
}

Let us now calculate $\bar\gamma$. The Lagrangian of a charge $q$ in a potential free condition is $L_0=\int T\ dt$. Now, an electrostatic potential $\Phi$ is turned on. The change in classical action $S$ is $\Delta S=\int L\ dt-\int L_0\ dt =\int(T-V)\ dt-\int T\ dt=- \int V\ dt=-q\int_0^t\Phi \ dt =-q \Phi\ t$. For an elementary electric charge $-e$ (so that $|q|>|e|$), we get $\Delta S=e\Phi t$. Now, if we define $\bar\gamma$ as the minimum unit of the electrostatic potential $\Phi$ allowed by the nature, corresponding to the minimum action $\hbar$, then  $\bar\gamma=\frac{\hbar}{e \cdot t}$. Therefore, the expression of the wave amplitude of the electrostatic field is given by $\psi_E={\bf e}^{i\frac{e \Phi t}{\hbar}}$ which has a form of ${\bf e}^{i\omega({\bf r})t}$ where $\omega({\bf r})$ is the angular frequency, dependent on the spatial coordinates. At a given location ${\bf r_0}$, it oscillates in time with a fixed frequency $\omega({\bf r_0})$. At a fixed time, the frequency becomes smaller, as $\Phi$ drops over distance. As expected, $\psi_E$ is not a travelling wave. When we talk about a charge $q$, the above expression changes to $\psi_E(q,\Phi)={\bf e}^{-i\frac{q\Phi t} {\hbar}}$.

Is $\psi_E={\bf e}^{i\frac{\Phi}{\bar\gamma}}$ still a solution, if the electrostatic field vanishes ($\mathcal E=0$), in which case
Eq.\eqref{eq7} reduces to $\nabla^2\psi_E=0$? We find that:
\begin{align}\label{eq7a}
    ln(\psi_E)&=i\frac{\Phi}{\bar\gamma}\nonumber\\
\implies\frac{\nabla\psi_E}{\psi_E}&=\frac{i}{\bar\gamma}\nabla\Phi\nonumber\\
\implies\frac{\nabla^2\psi_E}{\psi_E}-\frac{\nabla\psi_E\cdot\nabla\psi_E}{\psi_E^2}&=\frac{i}{\bar\gamma}\nabla^2\Phi
\end{align}
Thus, for $|\mathcal{E}|=|\nabla\Phi|=0$, ${\bf e}^{i\frac{\Phi} {\bar\gamma}}$ will be a solution of $\nabla^2\psi_E=0$, if
$\nabla\psi_E=0$. However, as we have already seen (in the footnote of pp. 5), the condition $\mathcal{E}=0$ automatically implies $\nabla \psi_E=0$. So, we can assert that the wave amplitude $\psi_E={\bf e}^{i\frac{\Phi}{\bar\gamma }}$ remains a valid wave amplitude of electrostatic field in a classically field free region. Therefore, the description may be relevant in the context of the Aharonov-Bohm effect.

\subsection{Magnetostatic case}
\subsubsection{Magnetostatic wave equation}
In the introduction, it was mentioned that if the free source current density $\vec{\mathcal{J}}_{free}={\bf{0}}$, one can conceive the principle: $\delta\int\mathcal{H}\ ds=0$. Hence, following the argument of section~\ref{Sec2}, we see that it is possible to devise the scalar wave equation:
\begin{align}\label{eq8}
\bar{\kappa}^2\nabla^2\psi_M+\mathcal{H}^2\psi_M&=0
\end{align}
where $\bar\kappa=\frac{\kappa}{2\pi}$ is a real number with the dimension of $\int\mathcal{H}\ ds$, i.e. that of electric current. Therefore, a simple solution of Eq.\eqref{eq8} can be given by $\psi_M={\bf e}^{i\frac{\int\mathcal{H}ds}{\bar\kappa}}$ (using the condition of non-existence of magnetic monopole: $\nabla \cdot\vec{\mathcal H}=0$). This wave amplitude can be meaningful when the value of $\bar\kappa$ cannot be neglected with respect to the dimension of $\int\mathcal{H}\ ds$ in the problem. 

Another possibility of a variational principle arises within the domain of magnetostatics, if the field $\vec{\mathcal{H}} \propto\nabla\times\vec{\mathcal A}={\bf0}$. In this case, we can conceive the variational principle $\delta\int\mathcal{A} \ ds=0$. A treatment similar to the one shown in section~\ref{Sec2} leads to:
\begin{align}\label{eq9}
\bar\eta^2\nabla^2\psi_A+\mathcal{A}^2\psi_A&=0
\end{align}
-where $\bar\eta=\frac{\eta}{2\pi}$ is a real number having the dimension of $\int\mathcal{A}\ ds$. In Eq.\eqref{eq9}, $\psi_A$ denotes the wave amplitude of the vector potential $\vec{\mathcal{A}}$ and is locally given by (under the Coulomb gauge condition $\nabla\cdot\vec{\mathcal A}=0$): $\psi_A={\bf e}^{i\frac{\int{\vec{\mathcal{A}}\cdot{d{\vec{s}}}}} {\bar\eta}}$. The description is relevant only when the factor $\bar\eta$ is non-negligible with respect to the dimension of $\int\mathcal{A}\ ds$ in the problem. The wave amplitude $\psi_A$ vanishes only in a special situation when $\vec{\mathcal A}$ diverges in a pathological condition. Before we proceed further, we note that the proposed form of solutions $\psi_M={\bf e}^{i \frac{\int\mathcal{H}ds}{\bar\kappa}}$ and $\psi_A={\bf e}^{i\frac{\int{\vec{\mathcal{A}}\cdot{d{\vec{s}}}}} {\bar\eta}}$ are not normalisable. One can, in principle, construct normalisable wave packets superposing several plane waves with appropriate weighting factors to explain 
general magnetostatic phenomena. But in the context of the magnetic Aharonov-Bohm effect where the magnetic field is zero, the non-normalisable wave amplitudes should be used.

The factor $\bar\eta$ can be calculated by keeping track of the change in action of an electric charge $q=-e$ placed in a electromagnetic vector potential. Now, when a charge $-e$ enters a magnetic field, its momentum changes from $m\vec{v}$ to $(m\vec{v}-e\vec{\mathcal{A}})$\footnotemark[2], where 
$\vec{\mathcal A}$ represents the electromagnetic vector potential. 

\footnotetext[2]{The convention used in the manuscript is consistent with the convention used in Feynman lectures – Volume III(21) The Schr$\rm\ddot{o}$dinger Equation in a Classical Context: A Seminar on Superconductivity [Eq.(21.15)], or in Volume II(15) – The Vector Potential. However, if we use the conventions used in the Advanced Quantum Mechanics by Sakurai [Eq. (1.84) at pp.15], then the linear momentum changes by $(\frac{e}{c})\vec {\mathcal{A}}$ where $c$ denotes the speed of light in vacuum. In this convention, if we set $c=1$ and note that $e$ is used as a negative constant in his book, we get back expression consistent with Feynman. 
} 

So, from the Maupertuis's principle, the change in action is given by $\Delta S=-e\int\vec{\mathcal A}\cdot{d\vec{s}}$. We can arrive the same result by considering the change in the Lagrangian $L\ \left(\frac{1}{2}mv^2-e\Phi\right) \rightarrow L'\ \left(\frac{1}{2}mv^2-e\Phi-e\vec{v}
\cdot\vec{\mathcal{A}}\right)$, as a charge $-e$ enters a region with vector potential $\vec{\mathcal A}$. Now, if we agree to denote the minimum allowed value of $\int\mathcal{A}\ ds$ as $\bar\eta$, corresponding to the minimum action $\hbar$, we find that: $\bar\eta=-\frac{\hbar}{e}$. This means that the wave amplitude of magnetic field takes the form $\psi_A({\bf r})={\bf e}^{-ie\frac{\int{\vec{\mathcal{A}}({\bf r})\cdot d\vec{s} }}{\hbar}}$. We note that this wave amplitude is also a non-travelling wave and is non-normalisable. It does not vanish, even when magnetic field $\vec{\mathcal{B}}$ is zero. For a charge $q$ ($|q|>|e|$), this phase will take the form: ${\bf e}^{iq\frac{\int{\vec{\mathcal{A}}({\bf r})\cdot d\vec{s}}}{ \hbar}}$. 

\section{Application to Aharonov-Bohm effect}\label{Sec3:AB}
\subsection{Influence of $\psi_E$ and $\psi_A$ on electron states}
Since the energy of a charge $q$ placed in an electric potential $\Phi$ increases by $W=q\Phi$ and the expression of wave amplitude $\psi_E$ is of the form of a time translation unitary operator $U_t={\bf e}^{-\frac{i}{\hbar}Wt}$, one can easily understand that the transformation of electron states in a potential $\Phi$ is given by $\psi_e\rightarrow \psi_e'=\psi_E({\bf r})\psi_e({\bf r})={\bf e}^{\frac{i}{\hbar} e \Phi({\bf r}) t}\psi_e$. So, the wave amplitude of the $field $ acts $locally$ on the electron state in such a way that it has traditionally been understood as the local unitary gauge transformation of electron state. The discussion in the previous section gives a new interpretation to the phase factor. The transformation $\psi_e'=\psi_E\psi_e$ can be assumed to be quantum version of the force $\vec{\mathcal {F}}$ experienced by a charge $q$ in an electric field $\vec{\mathcal{E}}$.

Similarly, as the linear momentum of a charge $q$ in an electromagnetic vector potential $\vec{\mathcal A}$ is given by $m\vec{v}+q\vec{\mathcal{A}}$ and the expression of wave amplitude $\psi_A$ is of the form of the spatial translation unitary operator $U_{\vec{a}}={\bf e}^{\frac{i}{\hbar}{\vec{p}} \cdot\vec{a}}$, we can make out that the transformation of the electron states in $\vec{\mathcal A}$ is given by $\psi_e\rightarrow\psi_e'=\psi_A({\bf r})\psi_e({\bf r})= {\bf e}^{-ie\frac{\int{\vec{\mathcal A({\bf r})}\cdot{d{\vec {s}}}}}{\hbar}}\psi_e$. Thus, the wave amplitude of the electromagnetic vector potential acts locally on the electron state in such a way that it has been interpreted as the gauge transformation of the electron state. While that understanding is perfect, this work presents a way to reinterpret the physical meaning of the phase factor.
In this case, however, the force interpretation does not make any sense, since the statement $\delta\int\mathcal A\ ds=0$ and Eq.\eqref{eq9} are valid only when magnetic field $\vec{\mathcal{B}}$ is zero and there is no magnetic force on the charge $q$. It is interesting to note that if a charge is placed in both the electrostatic and magnetostatic fields, then the transformation of the state $\psi_q$ of a charge $q$ is given by $\psi_q\rightarrow\psi_q'={\bf e}^{\frac{i}{\hbar}q\left[\int{\vec{\mathcal A}\cdot{d{\vec{s}}}}-\Phi t\right]}\psi_e$ which has the desirable travelling wave characteristic. 
\begin{figure}[ht]
\centering
\begin{subfigure}{0.5\textwidth}
\centering
    \includegraphics[width=6.0cm, height=4.0cm]{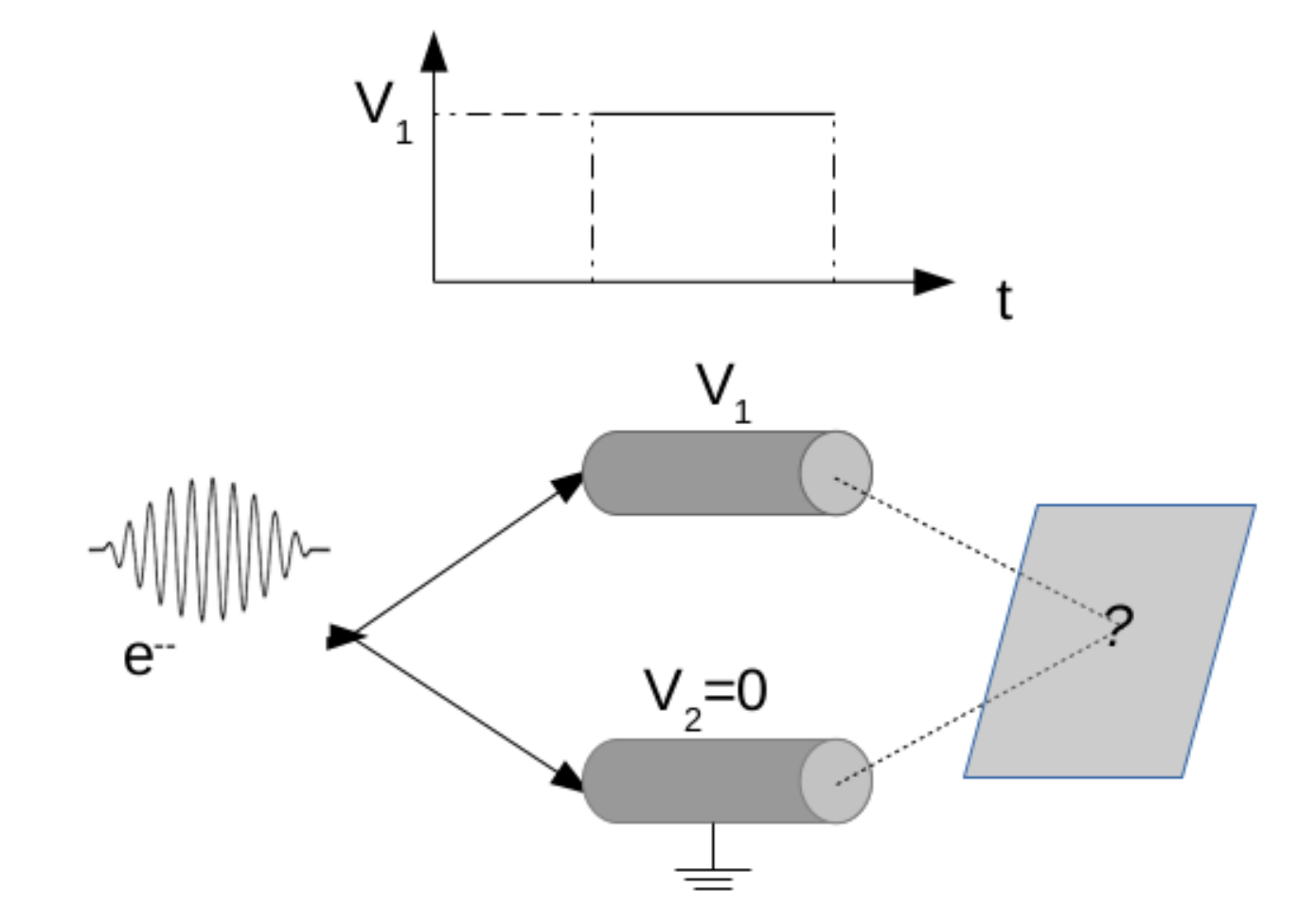}
    \caption{}
    \label{f1b}
\end{subfigure}
\hspace{-0.5cm}
\begin{subfigure}{0.5\textwidth}
    \includegraphics[width=7.0cm, height=4.0cm]{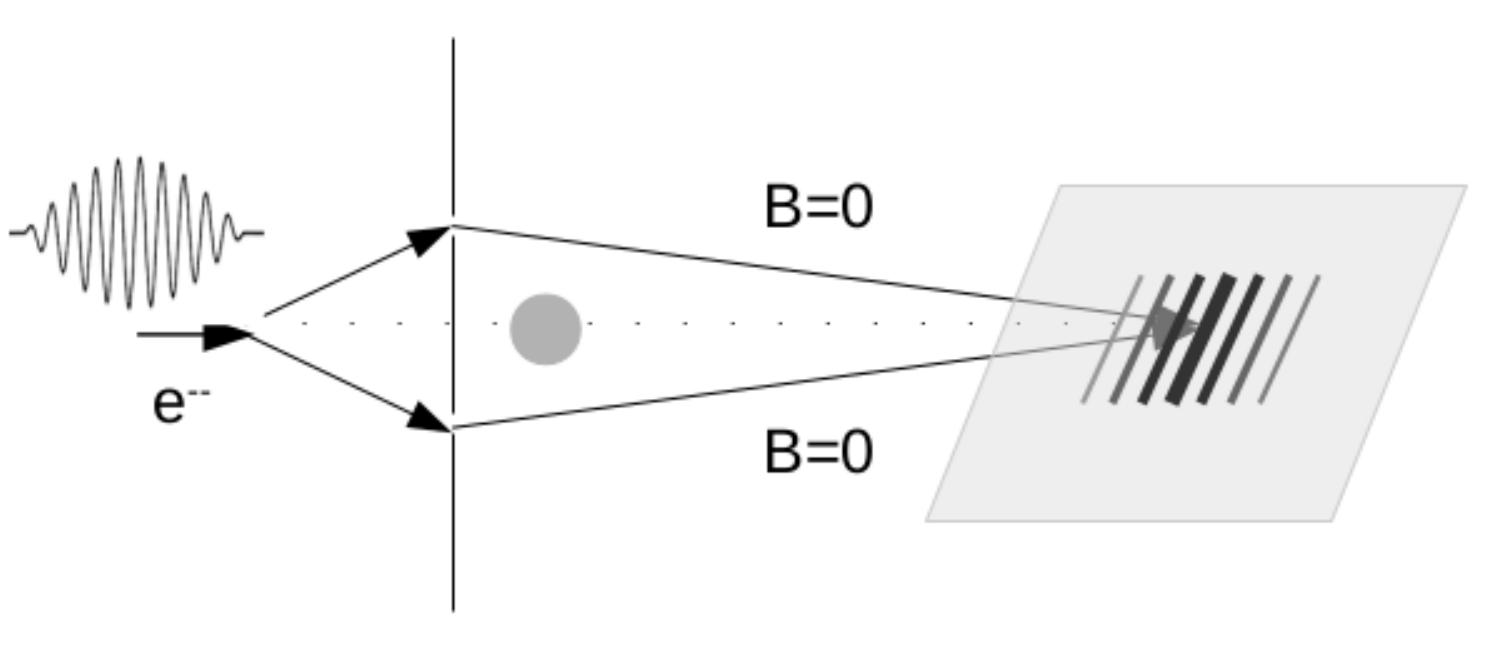}
    \caption{}
    \label{f1a}
\end{subfigure}%
\caption{Schematic diagrams of the experimental realisation of the (a) electrostatic Aharonov-Bohm effect and (b) magnetostatic Aharonov-Bohm effect.}
\label{fig}
\end{figure}
\vspace{-1.0cm}
\subsection{Explanation of the electrostatic Aharonov-Bohm effect}
The schematic diagram of the experiment is shown in the above Fig.~\ref{f1b}: the incident electrons are divided into two streams and are passed through the metal tubes maintained at constant but different potentials. The focus is to ensure that the electric field is exactly zero within these tubes. This is difficult to achieve for many reasons: first of all, it is practically impossible to ensure that fringing field is zero near the terminals of the tubes. Secondly, the electrons passing through the tube give rise to an electric field that induces a surface current~\cite{wang2015absence} on the inner wall of the tubes. This current could also generate electric field on the passing electrons. In real experiment, all these factors complicate the confirmation of the electrostatic Aharonov-Bohm effect.

Here we note that the electron wave functions along the two paths $^{(1)}$ and $^{(2)}$ transform as: $\psi_e^{(1)'}({\bf r})=\psi_E^{(1)}({\bf r}) \psi_e^{(1)}({\bf r})={\bf e}^{+i\frac{e \Phi^{(1)}\ t^{(1)}} {\hbar}}\psi_e^{(1)'}$ and $\psi_e^{(2)'}({\bf r})=\psi_E^{(2)}({\bf r})\psi_e^{(2)}({\bf r})={\bf e}^{+i \frac{e\Phi^{(2)}\ t^{(2)}}{\hbar}}\psi_e^{(2)'}$. Therefore,
the transformed electron wave functions could, in principle, interfere and lead to interference pattern. The key point is to ensure that the difference $|\Phi^{(1)}\ t^{(1)}-\Phi^{(2) }\ t^{(2)}|$ between two streams, including all experimental conditions should be of the same order of magnitude as $\frac {\hbar}{e}\sim10^{-15}$ V-s. If this quantity becomes $\gg10 ^{-15}$ V-s, then the wave amplitude $\psi_E$ of the electrostatic field will lose its relevance and classical electrostatics will predominate manifesting a vanishing electric field which suggests that there is no force on the electron.

\subsection{Explanation of the magnetic Aharonov-Bohm effect}
The experiment configuration of the magnetostatic Aharonov-Bohm effect is shown in the above figure~\ref{f1a}. It is tempting to consider Eq.\eqref{eq8} as a starting point. 
However, since the presence of the solenoid renders the space accessible to the electrons as a non-simply connected region, the concept of magnetic scalar potential cannot be used in this context. Therefore, the use of $\psi_M={\bf e}^{i\frac{\int\mathcal{H}\ ds} {\kappa}}$ appears to be improper. 

In this case, Eq.\eqref{eq8} reduces to $\nabla^2\psi_M=0 $. We also note that the condition of $\mathcal{H}=0$ signifies that $\nabla\psi_M=0$ (see the footnote in pp. 5). On the other hand, in this case the vector potential can be expressed as a gradient of a scalar potential: $\vec{\mathcal A}=-\nabla \mathcal{V}$ and it satisfies the Coulomb gauge condition $\nabla\cdot\vec{\mathcal{A}}=-\nabla^2\mathcal{V}=0$. We find that $\psi_M$ and $\mathcal{V}$, in the absence of magnetic field, are related in a way similar to $\psi_E$ and $\Phi$ in absence of electric field, which we already discussed in Eq.\eqref{eq7a}. Following that discussion, we see that the relevant wave amplitude in this case, can be given by ${\bf e}^{i\frac{\mathcal{V}}{\bar\beta}}={\bf e}^{i\frac{\int\vec{\mathcal{A}}({\bf r})\cdot d\vec{s}}{\bar\beta}}$, where $\bar\beta$ is a number with the dimension of $\int\mathcal{A} ds$. From the discussion following Eq.\eqref{eq9}, we can deduce that $\bar\beta$ and $\bar\eta$ are identical parameters.

This can be seen more directly in the following way. Though the electron does not come in contact of classical magnetic field, it does encounter the classical vector potential, so Eq.\eqref{eq9} applies. The electron wave function is transformed as: $\psi_e\rightarrow\psi_e'={\bf e}^ {i\frac{\int\vec{\mathcal A}({\bf r})\cdot d\vec{s}}{\bar \eta}}\psi_e$. Following our previous discussion, we can deduce that the relevant transformation of electron state in magnetic Aharonov-Bohm effect is $\psi_e\rightarrow \psi_e'=\psi_A({\bf r})\psi_e({\bf r})={\bf e}^{-ie\frac{\int {\vec{\mathcal A({\bf r})}\cdot{d{\vec {s}}}}}{\hbar}} \psi_e$. Then, the phase difference along two different electron paths $\mathcal{C}_1$ and $\mathcal{C}_2$ around the solenoid is proportional to: 

\begin{align}\label{eq30}
 \Delta{\mathcal{V}}&\propto\int_{\mathcal{C}_1}\vec{\mathcal A}\cdot{d{\vec{s}}}-\int_{\mathcal{C}_2}\vec{\mathcal A}\cdot{d{\vec{s}}}
 =\oint\vec{\mathcal A}\cdot{d{\vec{s}}}
 =\int_{\mathcal{S}}\nabla\times\vec{\mathcal{A}}\cdot{d{\vec{ S}}}
 =\Phi_B
\end{align}
Clearly, the quantity $\Phi_B=\oint\vec{\mathcal{A}}\cdot d{\vec{s}}$ must be of the same order of magnitude as the flux quantum $\frac{\hbar}{e}$ which we have identified as $\bar\eta$ in our formalism\footnotemark[3]. We note that the author of~\cite{kasunic2019magnetic} mentioned that the flux $\Phi_B$ should be of the order of $10^{-15}$ Wb which is also the same order of magnitude as $\frac{\hbar}{e}$, for the effect to manifest.
\footnotetext[3]{In Sakurai's notation, where $c$ is not set to 1, the expression for flux quantum would be $\frac{\hbar c}{e}$}.
If the magnetic field is non-zero in a region, then one can prove that the wave amplitude $\psi_M$ of the field acts on a particle with magnetic dipole moment $\vec{\mu} $ as: $\psi_{\mu}\rightarrow\psi'_{\mu}=\psi_M\psi_{\mu} ={\bf e}^{-i\frac{\mu\mathcal{B}t}{\hbar}}\psi_{\mu}$. A torque is exerted on the particle and the angular momentum comes from the magnetic field. However, if $\vec{\mu}={\bf0}$ or $\mathcal{B}={\bf0}$, a particle of charge $-e$ acquires a linear momentum that comes from the underlying electromagnetic vector potential. We saw that for $\mathcal{B}= {\bf0}$, $\psi_M$ reduces to $\psi_A={\bf e}^{-ie\frac{\int {\vec{\mathcal{A}({\bf r})}\cdot{d{\vec{s}}}}}{\hbar}}$. So, in the semi-classical description of magnetostatics, the wave amplitude of the magnetic field depends on the wave amplitude of the electromagnetic vector potential. This is different from the case in electrostatics. We conjecture the magnetic Aharonov-Bohm effect can also be considered as a local phenomenon although it is mediated by $\psi_A $, the wave amplitude of the vector potential (not that of the field), since (i) in this case $\psi_M\equiv\psi_A $ and (ii) $\psi_A$ carries the linear momentum to the regions where classical field vanishes.


\subsection{Aharonov-Casher effect}
Aharonov-Casher effect is a quantum mechanical phenomenon which asserts that a neutral particle with a non-zero magnetic dipole moment (fluxon, an infinitely thin solenoid or idealised line of flux~\cite{aharonovquantum}), like neutrons, should experience a phase shift when diffracted around a line of charge. This effect is dual to the Aharonov-Bohm effect and has been experimentally observed~\cite{cimmino1989observation}. In this case, the dipole moves under an electrostatic force field. Though there is a non-zero force here, the two effects are in fact the same phenomena in two dimension with a charge and a fluxon, only the reference frames are different. However, the wave amplitude $\psi_E$ cannot be expressed as $\psi_E={\bf e}^{-i\frac{q\Phi t}{\hbar}}$ in this electrostatic environment, because neutron does not possess electric charge and the concept of potential is not as relevant. So, there must be an alternative explanation. 

We note that a magnetic dipole moment $\vec{\mu}$ at rest in an electrostatic field carries a linear momentum even when it is not moving. This momentum is called `hidden momentum', and is expressed as ${\bf p}_{hid}=\vec{\mu}\times\vec {\mathcal E}$ when it experiences an electrostatic field $\vec{\mathcal {E}}$~\cite{kholmetskii2005hidden,griffiths2018catalogue}. It has been shown that for static fields, ${\bf p}_{hid}$ equals the negative of the electromagnetic momentum, i.e. ${\bf p}_ {EM}$~\cite{babson2009hidden}. But curl of ${\bf p}_{EM}$ is zero, because it can be expressed as a gradient of an action. It follows that the curl of the hidden momentum is zero as well. Therefore, the treatment of the section~\ref{Sec2} will be valid and the wave amplitude should have the form $\psi_{ \mu\mathcal{E}}={\bf e}^{i\frac {\int\vec{\mu}\times\vec{\mathcal{E}}\cdot d{\vec{s}}}{\bar \xi}}$, where $\bar\xi$ is a parameter like $\bar\gamma$ and the integral is taken along the local direction of the hidden momentum $\vec {\mu}\times \vec{\mathcal E}$. We can calculate the value of $\bar \xi$ in the following manner: the Lagrangian of a magnetic dipole $\vec{\mu}$ in the electric field $\vec{\mathcal E}$ is expressed as~\cite{anandan2000classical, kholmetskii2014electric}: $L=\frac {1}{2}mv^2+\vec{v}\cdot(\vec{\mu}\times\vec{\mathcal{E}})$. Now, clearly the introduction of the electric field changes action by $\Delta S=\int \vec{v}\cdot(\vec{\mu}\times\vec{\mathcal{E}}) dt =\int(\vec{\mu}\times\vec{\mathcal{E}})\cdot d\vec{s}$. Let us agree that $\bar\xi=\hbar$, the naturally allowed minimum unit of action. Then, the wave amplitude must be expressed as
$\psi_{\mu\mathcal{E}}={\bf e}^{i\frac{\int (\vec{\mu}\times\vec {\mathcal{E}})\cdot d\vec{s}}{\hbar}}$.
\subsection{He-McKellar-Wilkens Effect}
Another quantum mechanical phenomenon that is engendered due to the conception of the Aharonov-Bohm effect, is the He-McKellar-Wilkens effect which can be called the dual of the Aharonov-Casher effect. In this case, an electric dipole in the vicinity of the magnetic field is predicted to acquire a phase given by $\Phi_{HMW}=\frac{1}{\hbar}\oint(\mathcal{B}\times\vec{d})\cdot ds$~\cite{he1993topological,wilkens1994quantum}. We note that the vector field $(\vec{\mathcal{B}}\times\vec {d})$ is not a hidden momentum, but it comprises a part of the canonical momentum of an electric dipole moving in an electromagnetic field [Eq.(4)in~\cite{wilkens1994quantum}]. This momentum has also been termed as latent momentum in~\cite{kholmetskii2014electric}\footnotemark[4]. Under the conditions that the permanent dipole $\vec{d} $ does not change with time (i.e. $\dot{\vec{d}}={\bf0}$)
and that the electric field does not vary in the dipole direction, i.e. $(\vec{d}\cdot\nabla){\vec{\mathcal{E}}} =0$, Wilkens finds the equation of motion of the dipole [Eq(7) of~\cite{wilkens1994quantum}]:
\begin{equation*}
M{\ddot{\vec{R}}}=\vec{v}\times\nabla\times(\mathcal{B}\times\vec{d})
\end{equation*}
\footnotetext[4]{The same authors call $\vec{\mu}\times \vec{\mathcal{E}}$ as the magnetic hidden momentum and $\vec{\mathcal{B}}\times\vec{d}$ as the electric hidden momentum in~\cite{kholmetskii2016force}.}
Further, Wilkens argues that with an appropriately chosen magnetic field, the force and torque acting on dipole are zero, but the $\Phi_{HMW}$ is not. He brings the instance of a hypothetical system of infinite straight line of the magnetic monopoles with constant magnetic charge per unit length. The magnetic field due to this configuration has cylindrical symmetry and falls radially, i.e. $\mathcal{ B}\propto\frac{\hat{\bf e}_r}{r}$.

It is further shown that if the electric dipole $\vec{d}$ is parallel to the direction of the magnetic line charge, then this configuration yields $\nabla\times(\mathcal{B} \times\vec{d})\propto\delta(r)\vec{d}$- where $\delta(r)$ denotes the Dirac delta function, which suggests that the force acting on the dipole is zero except at $r=0$, i.e. at the location of the line charge. Thus, as long as the dipole is confined at $r>0$, the vector field $(\vec{\mathcal{B}}\times\vec{d})$ is curl-free and we can talk about the wave amplitude $\psi_{d\mathcal{B}}$ acting on it, which can be expressed as:
\begin{equation}
    \psi_{d\mathcal{B}}={\bf e}^{\frac{\int(\mathcal{B}\times\vec{d})\cdot ds} {\bar\chi}}
\end{equation}
Using the Lagrangian [Eq.(3) of~\cite{wilkens1994quantum}] of the dipole in magnetic field, we can easily establish that the factor $\bar\chi=\hbar$ if we use the convention $c=1$. With this, the effective wave amplitude acting on the dipole around the magnetic line charge can be shown to be equal to the He-McKellar-Wilkens phase itself.
\subsection{Quantum phases of the relativistic dipoles in electromagnetic fields}
The phases or wave amplitudes that appear in the context of the Aharonov-Bohm effect, Aharonov-Casher effect and the He-McKellar-Wilkens effect are not the only ones that could possibly emerge in the theory. In particular, one pertinent question is how do the quantum phases modify if these charges or dipoles move at the relativistic speeds. This may be relevant in the context of experiments trying to determine possible non-zero electric dipole moment of muon~\cite{bennett2009improved,abe2019new}. Kholmetskii et al.~\cite{kholmetskii2018quantum} derived these additional quantum phases, other than the ones responsible for the Aharonov-Casher effect and the He-McKellar-Wilkens effect, using the covariant Lagrangian for the dipole in an electromagnetic field (assuming $c=1$):
\begin{align}
    \delta_{d\mathcal{E}}&=\frac{1}{\hbar}\int\gamma_v(\vec{d}_{0||}\cdot\vec{\mathcal{E}})\vec{v}\cdot ds\nonumber\\
    \delta_{\mu\mathcal{B}}&=\frac{1}{\hbar}\int\gamma_v(\vec{\mu}_{0||}\cdot\vec{\mathcal{B}})\vec{v}\cdot ds
\end{align}
-where $\gamma_v=\frac{1}{\sqrt{1-(\frac{v}{c})^2}}$ and the subscripts $_0$ and $_{||}$ denote the component of the proper electric ($\vec{d}$) or magnetic ($\vec{\mu}$) dipole moments parallel to the velocity vector. They also derived the expressions of the phases of the charges, superposition of which might lead to the above expressions of the phases of the dipoles. The phases are dependent on the speed $v$ of the dipoles, since the quantum effects of the static fields, as seen by the dipoles, are modified by their speeds and the directions of motion. Kholmetskii, in an earlier paper, obtained the force on a dipole in an electromagnetic field [Eq(18)  of~\cite{kholmetskii2016force}]:
\begin{align}\label{DipoleForceEquation}
    \vec{F}&=\frac{d}{dt}\left(\gamma_vm{\vec{v}}\right)\nonumber\\
    &=\nabla({\vec{d}\cdot\vec{\mathcal{E}}})+\nabla({\vec{\mu}\cdot\vec{\mathcal{B}}})+\frac{d}{dt}\left(\gamma_v(\vec{d}_{0||}\cdot\vec{\mathcal{E}})\vec{v}\right)+\frac{d}{dt}\left(\gamma_v(\vec{\mu}_{0||}\cdot\vec{\mathcal{B}})\vec{v}\right)-\frac{d}{dt}(\vec{\mathcal{B}}\times{\vec{d}})-\frac{d}{dt}(\vec{\mu_0}\times{\vec{\mathcal{E}}})
\end{align}
-where we again used the convention that $c=1$. We have seen earlier that the phases in the context of the Aharonov-Casher effect and the He-McKellar-Wilkens effect could be explained using the semi-classical model of the static curl-free fields. The last two terms of Eq.\eqref{DipoleForceEquation} are `hidden momentum' terms that we found to be curl-free vector fields in the regions accessible to the dipoles. One may perhaps anticipate that the vector fields $\left(\gamma_v(\vec{d}_{0||}\cdot\vec {\mathcal{E}})\vec{v}\right)$ and $\left(\gamma_v(\vec{\mu }_{0||}\cdot\vec{\mathcal{B}})\vec{v}\right)$, which are momenta of dipoles, should be curl-free under appropriate conditions.
\subsection{Connection to vacuum fluctuations}
The preceding discussion inevitably brings the question forth whether the Aharonov-Bohm effect is an instance of the vacuum fluctuations. Though this idea appears very appealing, the current author thinks that this idea may be over-simplifying and needs to be scrutinised very carefully. Specifically, the electrostatic and the magnetostatic configurations prohibit one from attributing a harmonic oscillator description to the field. Therefore, `vacuum fluctuations' and zero point energy appear to be an ill-defined concept in this context. However, the situation has some similarity with vacuum fluctuations.

\section{Real world examples}\label{Sec4:RWE}
\subsection{I: \underline{Quantisation of electric charge}}
It is well-known that Dirac introduced magnetic monopoles~\cite{dirac1931quantised}, to explain the empirically known fact that electric charge is quantised. However, no magnetic monopole has been observed so far. Some authors attempted to address this question in the context of the magnetic Aharonov-Bohm effect~\cite{barone2005remark}. However, the invocation of the magnetic Aharonov-Bohm effect seems to be a desperate attempt to explain the quantisation. After all, this feature is known well within the context of electrostatics and there should be some explanation within this domain itself. In the following, we shall try to throw some light to this question.

Classically speaking, the electric field remains the same if the potential is changed by a constant. If we insist that the same should happen in quantum domain, then that means $\psi_E $ will remain the same if $\Phi$ is changed by a constant $\Phi_0$. This is usually referred to as gauge freedom when acting on a charge $q$.
\begin{align}\label{EqQuantisation0}
\psi_E(q,\Phi)&=\psi_E(q,\Phi+\Phi_0)\nonumber\\
\implies {\bf e}^{-i\frac{q\Phi t}{\hbar}} &= {\bf e}^{-i\frac{q(\Phi+\Phi_0) t}{\hbar}}
\end{align}
Collecting the cosine part of the phase, we find: 
\begin{align}\label{EqQuantisation2}
\cos\left(\frac{q(\Phi+\Phi_0)t}{\hbar}\right)&=\cos\left(\frac{q\Phi t}{\hbar} \right)\nonumber\\
\implies \frac{q(\Phi+\Phi_0)t}{\hbar}&=\frac{q\Phi t}{\hbar} + 2n\pi
\end{align}
-where $n$ is an integer [$n\in\mathbb{N}$]. From Eq.\eqref{EqQuantisation2}, it follows that
\begin{align}\label{EqChargeQuant}
    q\Phi_0 t&=n\ 2\pi\hbar=n\ 2\pi e\frac{\gamma}{2\pi}t\nonumber\\
\implies q&=ne\frac{\gamma}{\Phi_0}=\frac{n}{N}e
\end{align}
-where in the last equality, we have used the condition that $\Phi_0=N\gamma$. Eq.\eqref{EqChargeQuant} states that electric charge should be a rational multiple of the elementary charge of an electron. This is not the proof of quantisation of charge, but is consistent with
the observation that particles with fractional charges appear in the domains of particle physics (quarks) and in fractional quantum Hall effect. 

We note that in quantum domain, potential is no longer a continuous function and its minimum unit is $\gamma$. We comment that under the delicate experimental conditions in which potential changes bit by bit, due to transfer of the electrons one at a time, $N$ equals 1 corresponding to the minimum of $\Phi_0$ and in such cases, the quantised nature of $q$ is manifested. The examples are Millikan's oil drop experiment, or basic chemical reactions.

\subsection{II: Flux quantisation}
We consider the situation relevant in the context of magnetic Aharonov-Bohm effect where the magnetic field vanishes. The Hamiltonian formulation, in this case, has a phase space with momentum $\mathcal{A}_{x,y,z}$ along the vertical axis, and the coordinates $(x,y,z)$ along the horizontal axis. The unit area in this phase space is $\bar\eta=\oint\vec{\mathcal{A}} \cdot{d\vec s}=-\frac{\hbar}{e}$. Now, let us insist that the wave amplitude $\psi_A$ should not change when the electromagnetic vector potential is transformed by a constant $\vec{\mathcal {A}_0}$: $\vec{\mathcal{A}}\rightarrow{\vec{\mathcal{A}}+\vec {\mathcal{A}_0}}$:
\begin{equation*}
    {\bf e}^{-i\frac{e\int(\vec{\mathcal{A}}+\vec{\mathcal{A}_0})\cdot d{\vec{s}}}{\hbar}}={\bf e}^{-i\frac{e\int\vec{\mathcal{A}}\cdot d{\vec{s}}}{\hbar}}
\end{equation*}
Collecting the argument of the cosine part of both sides of the above equation, we find that:
\begin{align*}
    \left(\frac{e\int(\vec{\mathcal{A}}+\vec{\mathcal{A}_0})\cdot d{\vec{s}}}{\hbar}\right)&=\left(\frac{e\int\vec{\mathcal{A}}\cdot d{\vec{s}}}{\hbar}\right)+2k\pi\ ... ... ... ...\rm{where\ k\in\mathbb{N}}\nonumber\\
\implies e\int\vec{\mathcal{A}_0}\cdot d\vec{s}&=k\ 2\pi\frac{h}{2\pi}=k\ h\nonumber\\
\implies\int\vec{\mathcal{A}_0}\cdot d\vec{s}&=k\frac{h}{e}
\end{align*}
The above exercise demonstrates that the flux is quantised.

\section{Summary and Discussions}
In this paper, we demonstrated that a not-so-well-recognised variational principle of the curl-free vector fields can be used to develop a semi-classical non-relativistic theory of the electrostatic and the magnetostatic fields that can used to explain the non-locality problem of the Aharonov-Bohm effect. We showed that the wave amplitudes of the fields intrinsically have the form of a unitary phase even in the regions where classical field is zero. So, in Aharonov-Bohm effect, an effect similar to the vacuum fluctuations of the classical static fields is at play. The wave amplitudes act on the electron states in a fashion which has, hitherto, been interpreted as the gauge transformation of the electron wave function. Additionally, we made important observations about the quantisation of charge: that electric charges must occur as a rational multiple of the elementary charge of electrons. This manifests as the usual `quantisation of electric charge' in the experiments where $N=1$. One implication of this study is that there are no magnetic monopoles. Apart from this, we also saw its implication in the context of flux quantisation. Finally, we comment that these wave amplitudes of the curl-free static fields do not have polarisation, which is evident from the form of the wave equations, in that they bear striking similarity to the Klein-Gordon equation.

\section{Acknowledgements}
The author is very grateful to Prof. Debapriyo Syam for providing critical review about the meaning of the wave amplitudes; to Dr. Tanmay Das and Dr. Nur Jaman for valuable discussions; and to Dr. Tamali Sikder for her precious support. The valuable comments by the reviewers are also appreciated. Finally, this work is dedicated to the memory of my loving mother, Tapati Bhattacharya, who recently left us.
\section{References}
\bibliographystyle{unsrt}
\bibliography{AharonovBohm}
\end{document}